\documentclass{emulateapj}

\usepackage{natbib}
\begin{document}

\title{Salts and radiation products on the surface of Europa}
\author{M.E. Brown}
\affil{Division of Geological and Planetary Sciences, California Institute
of Technology, Pasadena, CA 91125}
\email{mbrown@caltech.edu}
\author{K.P. Hand}
\affil{Jet Propulsion Laboratory, California Institute of Technology, Pasadena,
CA 91109}

\begin{abstract}
The surface of Europa could contain the compositional imprint of a 
underlying interior ocean, but competing hypotheses differ over whether 
spectral observations from the Galileo spacecraft show the signature 
of ocean evaporates or simply surface radiation products unrelated to 
the interior. Using adaptive optics at the W.M. Keck Observatory, we 
have obtained spatially resolved spectra of most of the disk of Europa 
at a spectral resolution $\sim$~40 times higher than seen by the Galileo 
spacecraft. These spectra show a previously undetected
distinct signature of magnesium 
sulfate salts on Europa, but the magnesium sulfate is confined to
the trailing hemisphere and spatially correlated with the presence of
radiation products like sulfuric acid and SO$_2$.
On the leading, less irradiated, hemisphere, our observations 
rule out the presence of many of the proposed sulfate salts, 
but do show the presence of distorted water ice bands. 
Based on the association of the potential MgSO$_4$ detection
on the trailing side with other radiation products, we conclude
that MgSO$_4$ is also a radiation product, rather than a 
constituent of a Europa ocean brine. Based on ocean chemistry models,
we hypothesize that, prior 
to irradiation, magnesium is primarily in the form of MgCl$_2$, and we predict
that NaCl and KCl are even more abundant, and, in fact, dominate the
non-ice component of the leading hemisphere. We propose observational
tests of this new hypothesis.

\end{abstract}


\section{Introduction}
Jupiter's moon Europa may harbor a global salty liquid water ocean of
 $\sim$ 100 km in depth and 2-3 times the volume of all the liquid water 
on Earth \citep{Anderson1998, Kivelson2000, zimmer2000}. 
This liquid water, in combination with a rocky silicate sea floor and 
radiolytically produced surface oxidants, may provide for a 
chemically rich ocean that would be considered habitable by terrestrial standards 
\citep{Chyba2000, Hand2009a}. 
While the surface of Europa might contain important 
clues about the composition of an interior ocean, 
after almost a decade of scrutiny from the Galileo 
spacecraft, debate still persists about the nature of 
the surface chemistry and the relative roles of exogenous 
radiation processing versus endogenous oceanic emplacement. 
From the current data, Europa can be viewed as a purely passive 
ice shell onto which ion and electron bombardment create
a limited chemical cycle confined to a thin surface layer, 
or it can be seen as a geologically active body 
with a chemically rich ocean that leaves a compositional 
fingerprint on the surface ice.

One key to understanding the nature of Europa is determining 
whether the composition of the icy surface reflects the interior 
ocean chemistry.  Early spectroscopy results from the NIMS (Near 
Infrared Mapping Spectrograph) \citep{Carlson1992} 
instrument on the Galileo spacecraft suggested that the surface of 
Europa was dominated by hydrated sulfate salts of the sort that would 
be expected in evaporates from an internal ocean \citep{McCord1998a}. 
The finding that these spectral signatures are more prevalent in what appear 
to be younger terrains bolsters the hypothesis that these are 
more-recently emplaced evaporates \citep{McCord2001}. 
Further study of the spectra, however, showed that these same NIMS 
surface spectra could be explained equally well by a surface dominated 
by hydrated sulfuric acid \citep{Carlson1999a}. Sulfuric acid is an 
expected product of the bombardment of an icy surface with sulfur ions 
\citep{Carlson2002,Strazzulla2007}. At Europa these sulfur ions are ultimately 
derived from molecules such as SO$_2$ released from volcanoes on 
Io and subsequently dissociated, ionized, and accelerated by Jupiter's 
rapidly spinning magnetic field until they impact Europa. 
Such radiolysis could also explain the SO$_2$ \citep{Lane1981} 
and sulfur allotropes \citep{Carlson2009} seen on the surface of Europa 
and their preferential appearance on the more heavily bombarded 
trailing hemisphere \citep{Paranicas2001,Paranicas2002}. 
The presence of sulfuric 
acid and additional sulfur products on the surface of Europa 
appears nearly inescapable, yet the salt hypothesis also remains compelling.

Over the past decade, considerable work has gone into 
detailed modeling of individual surface units on Europa and developing 
a combined picture where hydrated sulfuric acid indeed dominates the 
trailing hemisphere, but hydrated salts of various compositions 
are more dominant elsewhere \citep{Dalton2005, Dalton2007, Shirley2010,
Dalton2012a, Dalton2012b}.
This analysis relies on attempting 
to fit the observed NIMS spectra by linearly combining spectral 
components from a library composed of laboratory spectra of hydrated 
salts, hydrated sulfuric acid, and water ice. Unfortunately, the 
NIMS spectra are at a sufficiently low spectral resolution that no 
distinct spectral features that would uniquely identify any of the 
non-ice spectral components are visible. Instead, the fit is largely 
determined by the broad shapes of the distorted water bands.
A careful examination of even the most salt-rich modeled spectra suggests 
that simple intimate mixtures of hydrated sulfuric acid and water 
ice are capable of providing satisfactory fits to the 
data \citep{Carlson2005}. While
both the hydrated salt and sulfuric acid hypotheses are compelling, 
we find, in agreement with \citet{McCord2010}, that the NIMS data provides too low of a spectral resolution 
to discriminate between these current hypotheses or provide alternatives.
Higher spectral resolution data has been obtained 
from ground-based telescopes, but, in general, these spectra cover an 
entire hemisphere of Europa, so icy and non-icy regions 
of the surface cannot be spatially discriminated \citep{Calvin1995}. 

Modern infrared spectrographs coupled with adaptive optics systems on 
large ground based telescopes have both the spatial and spectral 
resolution required to potentially answer the question of the 
composition of the non-ice material on Europa. 
An early observation of the trailing hemisphere of Europa using 
adaptive optics at the W.M. Keck observatory \citep{Spencer2006} 
showed no conclusive spectral features in the 1.50-1.75 $\mu$m wavelength 
region where many of the proposed salts should have spectral absorptions, 
suggesting that pure hydrated salts are not dominant on the trailing 
hemisphere, but with the limited spatial coverage and limited spectral
band, many additional explanations are possible. 

In order to continue the investigation of the composition of the surface of
Europa at higher spectral resolution, we obtained high 
resolution spatially resolved spectra of Europa in wavelength regions 
between 1.4 and 2.4 $\mu$m using the adaptive optics system and the 
OSIRIS integral field spectrograph at the W.M. Keck Observatory 
\citep{Larkin2003}. Here we report on these new observations,
discuss a new spectral feature discovered on the trailing hemisphere at
this resolution, and suggest a possible identification of the
species responsible and a scenario plausibly explaining the species and 
its spatial distribution.
\section{Observations and data reduction}
We observe Europa 
on the nights of 18, 19, and 
20 September 2011 (UT), covering nearly all Europa longitudes. 
In the setup used, OSIRIS obtained a 
simultaneous spectrum at each of 1024 spatially resolved points within 
a 0.56 by 2.24 arcsecond region of the sky at a pixel scale of 0.035 
by 0.035 acrseconds. At the time of observation, Europa had an angular 
diameter on the sky of 1.028 arcseconds, so the full disk of Europa could 
be covered in two pointings. We used both the broad band H and K band 
settings to cover wavelength ranges of 1.473 - 1.803 $\mu$m (Hbb setting) 
and 1.956 - 2.381 $\mu$m (Kbb setting), respectively. 
For each spectral setting we first observed half of Europa's disk, 
offset the telescope 20 arcseconds north to record a spectrum of the sky, 
and then offset back to the other half of Europa's disk. Solar type 
stars were observed at each spectral setting in order to calibrate 
telluric absorption and to construct a relative reflectance spectrum
(see Table 1 for observational details). 
The raw OSIRIS data were turned into a spectral data cube through the 
thoroughly automated OSIRIS DRP (data reduction package) that includes 
data extraction, calibration, and precise mosaicing. Because of the 
large range of air masses through which Europa and the calibrator stars 
were observed, we found that a single telluric absorption calibration 
correction was insufficient for both halves of a pair of Europa 
observations. We thus interrupted the DRP at the calibration step 
and manually corrected for telluric absorption. We first divided the 
spectrum by the spectrum of a solar-type telluric calibrator that was 
observed as close in airmass as possible. We further corrected these 
spectra by constructing a transmission function by dividing 
spectra of the same star taken at two different airmasses and 
dividing the spectrum of Europa by an exponentially scaled version 
of this transmission function. The scaling was found empirically 
for each individual observation. Because each observation included 
hundreds of individual spectra, the scaling and thus the transmission 
correction could be found to extremely high precision. 
In addition, we found that the DRP is not optimized for the extraction 
of precision spectra of continuum sources such as Europa. 
In particular, small scale spectral variations appeared that were 
correlated with location on the spectral format on each of the observations. 
To correct these, we 
implemented a separate calibration step where we combined spectra 
of a calibration star to fill the entrance aperture. We then extracted 
these spatially resolved spectra, which should be identical at each spatial position. 
We constructed a spatial-spectral flat-field map to normalize the 
spectra so they are identical. We then used this flat-field map to 
correct the Europa and the Europa calibration star spectra. After this 
additional step, no spatially correlated spectral features were 
observable in the extracted spectra of Europa. 
The final data product consists of a single data cube for each 
night at each setting (2 cubes were obtained for each setting 
on 19 September) with full wavelength 
and relative reflectance calibration of the spectrum at each point.
Over the three night period, 
we were able to obtain spectra at a spatial resolution of 
$\sim$150 km over nearly the entire surface of Europa 
at a spectral resolution approximately 40 times higher 
than that of the NIMS instrument on Galileo.

\section{Spectral maps}

As a first step to studying the surface composition, we generated a 
global map of the non-water ice material across the surface of Europa. 
Several model-dependent methods have been used in an attempt to map the 
abundance of the non-ice material on the surface of Europa 
\citep{McCord1999,  Carlson1999a, Grundy2007}.
We note that the general features of all of these methods is 
reproduced by making the simple observation that a 
distinguishing characteristic between the icy and non-icy material 
is the depth of the water ice absorption feature at 2 $\mu$m. 
We measure the depth of the 2$~\mu$m water absorption feature at each spatial 
point by taking a median of the reflectance between 2.198 $\mu$m and 2.245 $\mu$m 
and dividing by the median of the reflectance between 1.990 $\mu$m and 
2.040 $\mu$m. Maps of the value of this ratio are created 
for each of three nights of observation, and the maps are projected 
to a cylindrical projection using the mapping tools in the IDL 
software package. In regions of overlapping coverage between the 
three nights, the average value is used.
The resulting ratio images (Fig. 1), reproduce the basic features 
seen in previous maps. The low latitude trailing hemisphere 
contains the largest fraction of non-water ice material. 
The high latitude regions contain the most pure water ice. The low 
latitude leading hemisphere -- which had never been fully mapped before -- 
also contains a significant abundance of non-water-ice material, 
though the amount is less than on the trailing hemisphere, and 
the overall reflectance of this hemisphere is higher.
\begin{figure}
\plotone{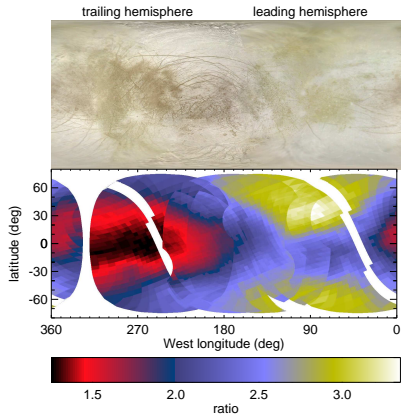}
\caption{The Voyager color base map of Europa compared to a 
cylindrically projected map of the ratio of the albedo at 2.2 $\mu$m to 
that at 2.0 $\mu$m, a proxy for the fractional contribution of 
water ice. This map reproduces the well-known result that 
the orbitally trailing side of Europa (centered at a longitude of 
270$^\circ$) contains the least pure water ice, while the 
higher latitude regions host the most pure ice. 
These data are the first to cover the complete leading hemisphere 
(centered at a longitude of 90$^\circ$), and show that 
this hemisphere, too, has less-pure ice than the poles.  }
\end{figure}

To study the composition of the surface materials on Europa, we 
extract spectra from two areas of the satellite: 
the dark low latitude trailing hemisphere with the 
highest fraction of non-water ice material 
(where the 2.2 to 2.0 $\mu$m ratio is less than 1.3)
and the bright low latitude 
leading hemisphere (with a ratio of less than 2.18) (Fig. 2). 
No calibration is available to turn these relative reflectance 
spectra into absolute reflectances. We estimate the 
absolute reflectances for each spectral data cube, however,
by taking a mean of all spectra from the surface of Europa 
and comparing to full-disk spectra of Europa previously calibrated by 
standard methods \citep{Calvin1995}. We anticipate that the 
absolute reflectance calibration is precise to approximately 10\%. 
The relative calibration between spatial locations and 
spectral settings should be significantly more precise.
\begin{figure}
\plotone{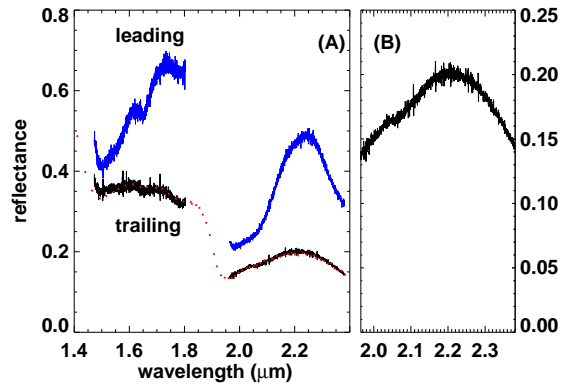}
\caption{(A) Representative reflectance spectra from the surface of 
Europa, including from the low latitude dark trailing side (black), 
the low latitude bright leading side (blue).
Uncertainty in the spectra 
can be seen from the scatter of the individual points. 
The red dots show the spectrum measured by NIMS from approximately
the same region as our trailing side spectrum \citep{Dalton2007},
showing the good agreement with our spectrum.
(B) An enlargement of the region of the trailing hemisphere spectrum 
showing the distinct 2.07 $\mu$m absorption feature. }
\end{figure}

The spectra are in good agreement with the lower resolution NIMS data 
obtained by the Galileo spacecraft on each of these general regions (Fig. 2), 
and, 
in general, at the higher resolution of the OSIRIS spectra 
little additional spectral detail is apparent. The most prominent 
exception, however, is the previously unobserved high 
resolution spectrum of the dark material on the trailing side of Europa. 
At the higher spectral resolution of our current data, 
the dark trailing side contains a clear spectral absorption 
feature at 2.07 $\mu$m that was not detectable by NIMS. 
The 2.07 $\mu$m absorption feature appears in the trailing
hemisphere equatorial spectra, 
but not in the simultaneously obtained polar spectra of the 
trailing hemisphere, confirming that the feature is real 
and not a function of the (identical) data reduction and calibration.

The appearance of the 2.07 $\mu$m feature on the trailing hemisphere 
and its non-appearance on the leading hemisphere suggests that
it might be related to the strong trailing hemisphere 
irradiation. To explore this hypothesis, we create a spatial map
of the strength of the 2.07 $\mu$m absorption.  We define the
continuum in the 2.07 $\mu$m region by fitting line segments to the 
region between 1.97 and 2.04 $\mu$m and between 2.10 and 2.18 $\mu$m
and taking the point closest to the data 
of the two line segments at each wavelength
as the continuum. We measure the strength of the absorption by
integrating the line absorption from 2.035 to 2.100 $\mu$m, and show
a spatial map of the absorption strength in Fig. 3. 
This method is crude but gives a simple continuum definition which
is continuous across different types of spectra ranging from concave up on the 
trailing hemisphere to concave down on the leading hemisphere.
The precise definition of this continuum
level is uncertain, so
the uncertainty in measuring the strength of this absorption
is moderately high, but the absorption feature nonetheless appears 
primarily confined to the trailing hemisphere (Fig. 1).   In fact, 
the 2.07 $\mu$m absorption feature appears well correlated
with the locations of the trailing hemisphere non-water ice component
and with the 
 UV absorption seen in Voyager data that has been attributed to the
presence of SO$_2$ -- a radiation product -- on the trailing hemisphere
\citep{Lane1981, mcewen1986}
It thus appears that the 2.07 $\mu$m absorption is strongly correlated
with the presence of SO$_2$, a clue that the species causing this
absorption is likely a radiolytic product. In the next section we attempt
to identify this product.
\begin{figure}
\plotone{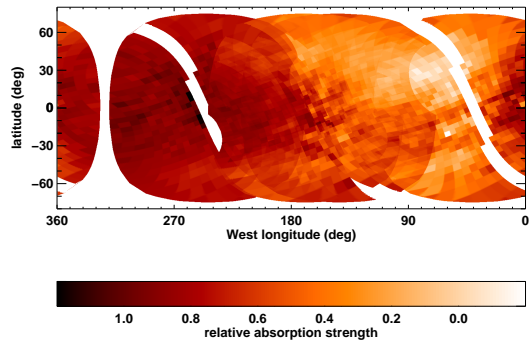}
\caption{A map of the strength of the 2.07 $\mu$m absorption. While
defining the continuum level across the feature is difficult, particularly 
across the transition from the non-icy trailing hemisphere to the icier
leading hemisphere, the map nonetheless suggests that the species
causing the 2.07 $\mu$m feature is predominantly located on the
trailing hemisphere, similar to known radiation products.}
\end{figure}

\section{Spectral identification and modeling}
We search proposed radiolytic products for 2.07 $\mu$m absorption 
features.  The primary products in the sulfur radiolytic cycle are 
sulfur allotropes, H$_2$SO$_4$, SO$_2$, S, and H$_2$S 
\citep{Carlson2002,Moore2007}. 
None of these has 
absorption features at or near 2.07 $\mu$m \citep{Carlson2002, fink1982}.
Expanding our search beyond already proposed radiation products, 
we examined cryogenic spectral libraries
\citep{Dalton2005, Dalton2007, Dalton2012}, 
products from ice irradiation experiments \citep[e.g.]{Hudson2001, Moore2007}, 
and the full ASTER \citep{Baldridge2009} and USGS \citep{Clark2007} 
digital spectral libraries (though as an important caveat, the spectra 
in these two libraries are all obtained at room temperature, and subtle 
features like that here are known to change with temperature). 
While minerals with features near 2.07 $\mu$m can occasionally be found,
in almost all cases there are stronger bands that would be predicted that
are not observed. 
Of all of the spectra examined, an appropriate 2.07 $\mu$m absorption
feature appears exclusively in some spectra of flash frozen 
MgSO$_4$ brines \citep{Dalton2005, Orlando2005} (brines, 
in this context, refers to saturated solutions which are flash frozen and
are thus not in a pure mineral form) 
and in measurements of the absorption coefficients of pure epsomite (MgSO$_4 \cdot$ 7H$_2$O) 
\citep{Dalton2012}. 

To expand our search, we obtained cryogenic laboratory spectra of a wide variety
of potentially plausible materials, including 
FeSO$_4\cdot$H$_2$O, FeSO$_4\cdot$7H$_2$O, MgS$_4$, 
MgSO$_4\cdot$7H2O, CaSO$_4\cdot$0.5H$_2$O, H$_2$SO$_4$, 
(NH$_4$)2SO$_4$, Na$_2$CO$_3$, MgCO$_3$, NH$_4$Cl, NaCl, 
KCl, NaSCN, NaOCl, Mg(OH)$_2$, NaOH, H$_2$O$_2$, CH$_3$OH, 
C$_2$H$_5$OH, NH$_3$, and CO$_2$. 
Samples were placed on gold target in small liquid nitrogen dewar 
with a nitrogen purge and spectra were collected with an 
Analytical Spectral Devices FieldSpec spectroradiometer covering the 
range from 0.35-2.5 $\mu$m, and with a MIDAC M4500 Fourier transform 
spectrometer covering the range from approximately 1.5-16 $\mu$m. 
Salts were examined in pure crystalline form and after dissolution 
in ultrapure deionized water. Additionally, salts and their 
crystalline brine forms were crushed, sieved and analyzed at 77 K with 
the above spectrometers for grain size fractions of $<63$ $\mu$m, 
63-106 $\mu$m, 106-180 $\mu$m, 180-300 $\mu$m, and 300-600 $\mu$m.
Example spectra are shown in Fig. 4.
Most spectra show no absorption features near the observed 2.07 $\mu$m
region. 
The sole exception is epsomite (MgSO$_4 \cdot $7 H$_2$O), where we 
also detect the same 2.07 $\mu$m
absorption feature seen in magnesium sulfate brines \citep{Dalton2005,
Orlando2005} and in previous laboratory spectra of epsomite \citep{Dalton2012}.
\begin{figure}
\plotone{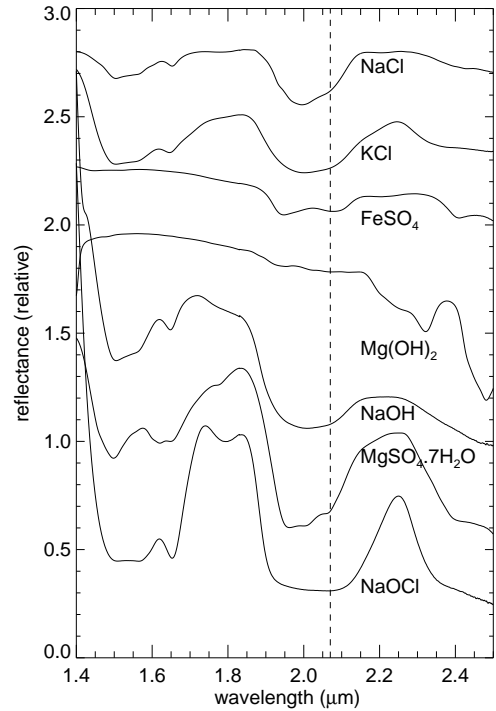}
\caption{Example cryogenic laboratory spectra of frozen crushed brines
and of epsomite. 
The spectra are 
scaled to unity at 1.8 $\mu$m and each spectrum is offset by 
0.3 units from the spectrum below it. The location of the 2.07 $\mu$m
absorption feature seen on the trailing hemisphere of Europa is 
shown for comparison.
In these and all other spectra obtained,
a 2.07 $\mu$m absorption feature was seen only for pure epsomite
(MgSO$_4\cdot$7H$_2$O).}
\end{figure}

Magnesium sulfate brine, water ice, and hydrated sulfuric acid are 
components of the Dalton et al. 
spectral libraries, so we experiment with the linear combination 
analysis as performed in \citet{Dalton2007} and subsequent papers.
As a check of the spectral library and the method, we first model all of
the spectra from
\citet{Dalton2007} and \citet{Shirley2010} and find that we recover the same
fractional mineral abundances within a few percent. We next turn to modeling
our spectrum of the equatorial region of the trailing hemisphere. Our
$\chi^2$ minimization finds that the best fit to this 
spectrum is composed nearly
exclusively of sulfuric acid hydrate (97\%) with a small amount (3\%) of 
100$\mu$m grain water ice (Fig. 5). The model fit is only moderately
good, and, critically, does not contain any of 
the magnesium sulfate brine required
to cause a 2.07 $\mu$m absorption feature. We can only match the 
2.07 $\mu$m feature 
by fixing the magnesium sulfate brine abundance to be 
about 30\%. A spectral model
with 34\% sulfuric acid hydrate, 30\% magnesium brine, 
and 36\% dark neutral material
provides an excellent fit to the 2 - 2.4 $\mu$m section 
of the spectrum -- including fitting the 
location, width, and depth of the 2.07$\mu$m feature 
-- but the fit to the continuum level of
the 1.4 - 1.8 $\mu$m data is poor.  The incorrect continuum fit 
to the 1.4 - 1.8 $\mu$m could be explained by at least one missing component
in the spectral library. If the library contained a 
component that was mostly smooth across the 
two spectral bands but was more reflective than sulfuric 
acid hydrate at 1.5 $\mu$m, the
algorithm would have selected this material to 
better match the spectrum. We discuss possibilities
for such a material later.
\begin{figure}
\plotone{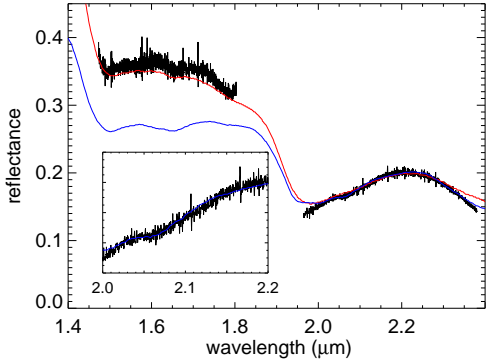}
\caption{The spectrum of the low-latitude trailing hemisphere.
The best linear spectral model consists almost entirely of
hydrated sulfuric acid (red), but this model has no absorption at
2.07 $\mu$m absorption. A good fit to the 2.0-2.4 $\mu$m portion
of the spectrum can be made by including nearly equal amounts
of hydrated sulfuric acid and magnesium sulfate brine along with
a dark neutral material (blue). While the fit of the 2.07 $\mu$m
absorption feature is excellent, the fit to the continuum between 
1.4 and 1.8 $\mu$m is poor. The inability to fit the full spectral
region demonstrates the inadequacy of both the spectral library used
and the general approach of linear spectral modeling.}

\end{figure}

The good spectral match of the 2.07 $\mu$m absorption 
feature to magnesium sulfate brines
(and to epsomite) -- along with the failed search 
to find any other material with a similar
absorption feature -- is compelling evidence that the new feature 
is indeed due to 
magnesium sulfate on the surface of Europa. 
While it is never possible to exhaust
every other possibility, no other plausible species 
matches the spectrum.

\section{Search for additional species}
In the analyses of Dalton and collaborators
\citep{Dalton2007, Shirley2010, Dalton2012a, Dalton2012b}, 
a general case is made
that sulfate salts are ubiquitous on the surface of Europa but that
the trailing hemisphere is dominated by radiolytically 
produced H$_2$SO$_4$. The mixing ratio of salt to acid is expected to 
increase from the trailing to the leading hemispheres. Again,
however, the NIMS data show no distinctive absorption features of
sulfate salts, but the overall spectral shape can be matched by
various linear combinations of these salts.

The higher resolution spectra available here allow us to qualitatively
examine this model and search for positive spectral evidence of these
proposed materials. To study the region of the surface expected to
be most salt-rich, we examine 
the spectrum of the leading hemisphere (Fig. 6). We
attempt linear spectral modeling using the Dalton et al.
library. Least-squares fitting finds what appears
to be a moderately good solution with 35.4\% sulfuric acid hydrate,
18.2\% hexahydrite, 17.9\% mirabilite, 7.7\% 100 $\mu$m grains of
ice, and 20.6\% 250 $\mu$m grains of ice. 

At the spectral resolution 
of the NIMS data,
this spectral fit would be considered excellent. The overall levels 
and shapes of
the spectrum are nicely matched. 
Our spectra, however, show no evidence for many of the subtle
absorption features that should be seen at the higher spectral
resolution of our data. Hexahydrite and mirabilite both cause a
small unobserved absorption feature at 1.6 $\mu$m, while mirabilite has
a 2.18 $\mu$m absorption. Indeed, the spectrum of the leading hemisphere
is generally quite smooth with no clear spectral features other than
those due to water ice.
\begin{figure}
\plotone{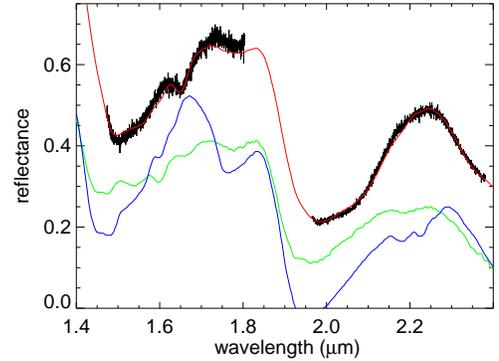}
\caption{The low-latitude leading hemisphere spectrum of Europa compared
to the best-fit linear model (red), which consists of  
35.4\% sulfuric acid hydrate,
18.2\% hexahydrite, 17.9\% mirabilite, 7.7\% 100 $\mu$m grains of
ice, and 20.6\% 250 $\mu$m grains of ice.  While the spectral fit appears
good, careful inspection reveals that hexahydrite (green) and 
mirabilite (blue) have distinct absorption features that should be
detectable in the spectrum. Hexahydrite and mirabilite both cause a
small unobserved absorption feature at 1.6 $\mu$, while mirabilite has
a 2.18 $\mu$ absorption. The spectrum of the leading hemisphere
is generally quite smooth with no clear spectral features other than
those to to water ice.}
\end{figure}

We conclude that we can find no evidence,
on regional scales, 
of the specific salts modeled by Dalton et al. 
The apparent requirement for the presence of these 
species in the linear spectral modeling is simply a function of
the limited number of species 
considered in the spectral library and the lack of distinct
discriminating spectral features that can be seen at
low resolution. While the spectral modeling
appears to be indicating that some non-water ice component, non-sulfuric
acid component is increasing in abundance from the trailing to the leading
hemisphere, the current spectral libraries being used do not appear to 
contain the species that is causing this behavior. 

\section{Discussion: the sea salt hypothesis}
The lack of any clear detection of
 sulfate salts on the leading hemisphere, coupled
with the possible detection of magnesium sulfate salts on the trailing
side is inconsistent with any of the hypotheses currently considered
for the composition of the surface of Europa. Assuming that the
2.07 $\mu$m absorption is indeed due to magnesium sulfate,
we consider an alternative hypothesis which could explain
the observations.

The spatial correlation of magnesium sulfate with the radiolytically-produced
sulfuric acid suggests the potential of an analogous magnesium
radiation cycle. If magnesium is present on the surface of Europa as
MgX, where is X is an atomic or radical anion, bombardment by sulfur ions, in the presence of water, could
yield a reaction such as
$$MgX_{n} + 4H_2O + S^+ \rightarrow MgSO_4 + 4H^+ + nX^-.$$

Such a cycle would suggest that MgX$_n$ is a pre-radiation component
of the surface of Europa. 
The presence of Na and K in the sputtered atmosphere of
Europa \citep{Brown1996, Brown2001} suggests that these cations 
should also be present, and that NaX and KX are likely also emplaced
on the surface. 

Sulfate has long been considered a plausible component
of a Europa ocean based on both theoretical \citep{Kargel1991, Kargel2000} and
experimental \citep{Fanale2001} studies, though more detailed
evolutionary models have suggested that sulfate 
could be much less abundant than initially expected \citep{McKinnon2003}.
In light of the lack of spectral evidence for sulfates on the leading
hemisphere of Europa and keeping in mind the theoretical difficulties
of high sulfate abundances, we explore alternative ocean chemistries.
In the reduced ocean models of \citet{Zolotov2008}, for example,
the most abundant cations are
Na and K, the most abundant anion is chlorine. 
We thus consider
the possibility here that the most abundant salts in Europa
ocean brines are not sulfates, but are chlorides. 

We consider the following 
conceptual model. Before irradiation, the non-water ice component
of the surface of Europa
is dominated by sodium and potassium chlorides, with magnesium
chloride a more minor component. With sulfur bombardment, these 
chlorides are radiolytically converted to sulfates. 
Sodium and potassium sulfates
are more easily sputtered than magnesium 
sulfates \citep{McCord2001}, so
the trailing hemisphere shows an enhancement in magnesium sulfate 
above the initial magnesium mixing ratio. Nonetheless sodium
and potassium sulfates should be present on the trailing side
with the magnesium sulfates. Indeed, the spectrum of
brines of sodium sulfate \citep{Mccord2002, Orlando2005, Dalton2005} have the 
general characteristic needed to better fit the continuum level
on the trailing hemisphere.

The leading hemisphere should contain the spectral signature of
NaCl and KCl salts (with a smaller amount of MgCl$_2$).
NaCl, KCl, and MgCl$_2$ are all spectrally flat and, when hydrated,
give distorted water bands with no distinct spectral features (see Fig. 4). 
Linear spectral models can easily be created which fit the 
leading hemisphere spectrum with these, rather than sulfate, salts,
but we regard this fact more as a demonstration that linear spectral
modeling, when faced with a spectrum that appears to be some sort
of distorted water-ice spectrum, and when given a library of various
distorted water-ice spectra, can nearly always find an adequate
fit to the large-scale features of the spectrum. Evidence for
the spectral detection of any salt materials on Europa should probably
only be considered valid when a distinct spectral feature caused
by the species can be specifically detected.

Because of their lack of distinct features in reflectance
spectra, these chloride
salts will prove much more difficult to spectrally
rule out (or confirm) than the sulfate salts were.
Like the detections of Na and K, it might prove more
feasible to detect the presence of these compounds in gas
phase once they have been sputtered from the surface, instead. 
Chlorine ions are detectable in the Io plasma torus
\citep{Kuppers2000,Feldman2001} and atomic chlorine
has been detected in Io's atmosphere \citep{Feaga2004},
but no search in the vicinity of Europa has been reported.
NaCl and KCl vapor have been detected at millimeter 
wavelengths in circumstellar environments \citep{Cernicharo1987}. 
The feasibility of observations such as these should be studied to
determine if there are possible ways to test the hypothesis
that chloride salts dominate the non-irradiated non-water ice 
component of the surface of Europa.

\section{Conclusion}
Spatially resolved high resolution spectroscopy of the surface of
Europa has allowed us to detect a previously unknown
absorption feature at 2.07$\mu$m on the surface of Europa. This 
absorption feature is spatially correlated with other radiation 
products on the trailing hemisphere of Europa and is thus likely 
radiolytically produced. The only plausible spectral match to 
this absorption feature that we find is to the spectrum of
MgSO$_4$ brine or that of the mineral epsomimte (MgSO$_4\cdot$7H$_2$O).

An examination of the spectrum of the less-irradiated leading hemisphere
reveals that, while distorted water ice bands are indeed present
at low latitudes, no evidence can be found that these distorted
bands are caused by sulfate salts. 
Inclusion of these salts in models for the leading hemisphere spectrum
results in distinct absorption features which are not observed in
the high resolution spectra.

Based on the association of the potential MgSO$_4$ detection
on the trailing side with other radiation products, we conclude
that MgSO$_4$ is also an irradiation product, rather than a 
constituent of a Europa ocean brine. We hypothesize that, prior 
to irradiation, magnesium is primarily in the form of MgCl$_2$, and we predict
that NaCl and KCl are even more abundant, and, in fact, dominate the
non-ice component of the leading hemisphere. These salts are
difficult to confirm spectroscopically in solid state form, but we
suggest that detection of chlorine and NaCl in ionized or atomic
state after sputtering might be possible.

\acknowledgements KPH acknowledges support from the Jet Propulsion 
Laboratory, California Institute of Technology, under a contract with 
the National Aeronautics and Space Administration and funded in part 
through the internal Research and Technology Development program. 
and from the NASA Astrobiology Institute, 
through the 'Astrobiology of Icy Worlds' node at JPL. 
MEB is supported by the Richard and Barbara Rosenberg Professorship 
at the California Institute of Technology.

\begin{deluxetable}{lllllll}

\tablehead{\colhead{Time}& \colhead{target}&\colhead{spectra}&\colhead{int.}&\colhead{airmass}& \colhead{Europa}&\colhead{comments}\\
\colhead{ (UT)} &\colhead{} &\colhead{setting} & \colhead{time}  & \colhead{}&\colhead{ lon.}  & \colhead{}\\
\colhead{}     &\colhead{} & \colhead{}       &\colhead{ (sec)} &\colhead{} & 
\colhead{(deg)} &\colhead{}}
\startdata
18 Sep 2011					\\
09:10&Europa, west&	Kbb&	900&	1.80&	41& \\
09:26&Europa, east&	Kbb&	900&	1.64&	43& \\	
09:44&sky&	Kbb&	450&	1.50&		\\
10:07&NLTT7066&	Kbb&	100&	1.37&\ &		telluric calibrator\\
10:09&NLTT7066&	Kbb&	100&	1.35&&		telluric calibrator\\
10:12&NLTT7066&	Hbb&	50&	1.34&&		telluric calibrator\\
10:14&NLTT7066&	Hbb&	50&	1.33&&		telluric calibrator\\
10:22&Europa, west&	Hbb&	450&	1.29&	47&	\\
10:31&Europa, east&	Hbb&	450&	1.25&	48&	\\
10:40&sky&	Hbb&	450&	1.22&&		\\
\\
19 Sep 2011	\\				
08:48&HD9986&	Kbb&	20&	1.50&&		telluric calibrator\\
08:49&HD9986&	Kbb&	20&	1.49&&		telluric calibrator\\
08:53&NLTT7066&	Kbb&	20&	1.91&&		transmission calibrator\\
08:54&NLTT7066&	Kbb&	20&	1.90&&		transmission calibrator\\
09:05&Europa, west&	Kbb&	900&	1.81&	143&	\\
09:21&sky&	Kbb&	900&	1.65&&		\\
09:37&Europa, east&	Kbb&	900&	1.51&	145&	\\
09:57&NLTT7066&	Kbb&	20&	1.40&&		transmission calibrator\\
09:58&NLTT7066&	Kbb&	20&	1.39&&		transmission calibrator\\
10:05&HD9986&	Hbb&	20&	1.16&&		telluric \&  transmission calibrator \\
10:06&HD9986&	Hbb&	20&	1.16&&		\\
10:48&Europa, west&	Hbb&	450&	1.18&	150&	\\
10:57&sky&	Hbb&	450&	1.15&	\\
11:06&Europa, east&	Hbb&	450&	1.13&	151&	\\
13:58&HD9986&	Kbb&	20&	1.12&&		telluric calibrator star\\
13:59&HD9986&	Kbb&	20&	1.13&&		telluric calibrator star\\
14:04&Europa, west&	Kbb&	900&	1.04&	164&	\\
14:20&sky&	Kbb&	900&	1.06&&		\\
14:37&Europa, east&	Kbb&	900&	1.09&	166&	\\
14:58&HD9986&	Hbb&	20&	1.33&&		telluric \& transmission calibrator\\
14:59&HD9986&	Hbb&	20&	1.34&&		\\
15:05&Europa, west&	Hbb&	450&	1.16&		\\
15:13&sky&	Hbb&	450&	1.18&		\\
15:22&Europa, east&	Hbb&	450&	1.21&		\\
\\
20 Sep 2011					\\
09:09&Europa, west&	Kbb&	900&	1.72&	244&	\\
09:26&sky&	Kbb&	900&	1.57&		\\
09:43&Europa, east&	Kbb&	900&	1.45&	247&	\\
10:03&NLTT7066&	Kbb&	20&	1.35&&		telluric calibrator star\\
10:04&NLTT7066&	Kbb&	20&	1.34&&		telluric calibrator star\\
10:19&HD9986&	Hbb&	20&	1.12&&		telluric calibrator star\\
10:21&HD9986&	Hbb&	20&	1.11&&		telluric calibrator star\\
10:33&Europa, west&	Hbb&	450&	1.22&	250&	\\
10:42&sky&	Hbb&	450&	1.18&		\\
10:51&Europa, east&	Hbb&	450&	1.16&	252&	\\
\enddata
\end{deluxetable}


\clearpage

\begin{thebibliography}{47}
\expandafter\ifx\csname natexlab\endcsname\relax\def\natexlab#1{#1}\fi

\bibitem[{Anderson {et~al.}(1998)Anderson, Schubert, Jacobson, Lau, Moore, \&
  Sjogren}]{Anderson1998}
Anderson, J.~D., Schubert, G., Jacobson, R.~A., Lau, E.~L., Moore, W.~B., \&
  Sjogren, W.~L. 1998, Science, 281, 2019

\bibitem[{Baldridge {et~al.}(2009)Baldridge, Hook, Grove, \&
  Rivera}]{Baldridge2009}
Baldridge, A., Hook, S., Grove, C., \& Rivera, G. 2009, Remote Sensing of the
  Environment, 113, 711

\bibitem[{Brown(2001)}]{Brown2001}
Brown, M.~E. 2001, Icarus, 151, 190

\bibitem[{Brown \& Hill(1996)}]{Brown1996}
Brown, M.~E. \& Hill, R.~E. 1996, Nature, 380, 229

\bibitem[{Calvin {et~al.}(1995)Calvin, Clark, Brown, \& Spencer}]{Calvin1995}
Calvin, W.~M., Clark, R.~N., Brown, R.~H., \& Spencer, J.~R. 1995, Journal of
  Geophysical Research-Planets, 100, 19041

\bibitem[{Carlson {et~al.}(2002)Carlson, Anderson, Johnson, Schulman, \&
  Yavrouian}]{Carlson2002}
Carlson, R.~W., Anderson, M.~S., Johnson, R.~E., Schulman, M.~B., \& Yavrouian,
  A.~H. 2002, Icarus, 157, 456

\bibitem[{Carlson {et~al.}(2005)Carlson, Anderson, Mehlman, \&
  Johnson}]{Carlson2005}
Carlson, R.~W., Anderson, M.~S., Mehlman, R., \& Johnson, R.~E. 2005, Icarus,
  177, 461

\bibitem[{Carlson {et~al.}(2009)Carlson, Calvin, Dalton, Hansen, Hudson,
  Johnson, McCord, Moore, McKinnon, \& Khurana}]{Carlson2009}
Carlson, R.~W., Calvin, W.~M., Dalton, J.~B., Hansen, G.~B., Hudson, R.~L.,
  Johnson, R.~E., McCord, T.~B., Moore, M.~H., McKinnon, W.~B., \& Khurana,
  K.~K. in Europa, ed. R.~T. Pappalardo (Tucson, AZ: Univ. Arizona Press), 283

\bibitem[{Carlson {et~al.}(1999)Carlson, Johnson, \& Anderson}]{Carlson1999a}
Carlson, R.~W., Johnson, R.~E., \& Anderson, M.~S. 1999, Science, 286, 97

\bibitem[{Carlson {et~al.}(1992)Carlson, Weissman, Smythe, \&
  Mahoney}]{Carlson1992}
Carlson, R.~W., Weissman, P.~R., Smythe, W.~D., \& Mahoney, J.~C. 1992, Space
  Science Reviews, 60, 457

\bibitem[{{Cernicharo} \& {Guelin}(1987)}]{Cernicharo1987}
{Cernicharo}, J. \& {Guelin}, M. 1987, \aap, 183, L10

\bibitem[{Chyba(2000)}]{Chyba2000}
Chyba, C.~F. 2000, Nature, 403, 381

\bibitem[{Clark {et~al.}(2007)Clark, Swayze, Wise, Livo, Hoefen, Kokaly, \&
  Sutley}]{Clark2007}
Clark, R., Swayze, G., Wise, R., Livo, E., Hoefen, T., Kokaly, R., \& Sutley,
  S. 2007, USGS digitial spectral library splib06a: U.S. Geological Survey, Digital Data Series 231.

\bibitem[{Dalton \& Pitman(2012)}]{Dalton2012}
Dalton, J.~B., \& Pitman, K.~M. 2012, Journal of Geophysical Research
  (Planets), 117, 09001

\bibitem[{Dalton(2007)}]{Dalton2007}
Dalton, J.~B. 2007, Geophysical Research Letters, L21205, doi:10.1029/2007GL031497, 34

\bibitem[{Dalton {et~al.}(2005)Dalton, Prieto-Ballesteros, Kargel, Jamieson,
  Jolivet, \& Quinn}]{Dalton2005}
Dalton, J.~B., Prieto-Ballesteros, O., Kargel, J.~S., Jamieson, C.~S., Jolivet,
  J., \& Quinn, R. 2005, Icarus, 177, 472

\bibitem[{Dalton {et~al.}(2012{\natexlab{a}})Dalton, Shirley, \&
  Kamp}]{Dalton2012a}
Dalton, J.~B., Shirley, J.~H., \& Kamp, L.~W. 2012{\natexlab{a}}, Journal of
  Geophysical Research-Planets, E03003, doi:10.1029/2011JE003909, 117

\bibitem[{Dalton {et~al.}(2012{\natexlab{b}})Dalton, Cassidy, Paranicas,
  Shirley, Prockter, \& Kamp}]{Dalton2012b}
Dalton, J.~I., Cassidy, T., Paranicas, C., Shirley, J., Prockter, L.~M., \&
  Kamp, L. 2012{\natexlab{b}}, Planetary and Space Science, in press

\bibitem[{Fanale {et~al.}(2001)Fanale, Li, De~Carlo, Farley, Sharma, Horton, \&
  Granahan}]{Fanale2001}
Fanale, F.~P., Li, Y.~H., De~Carlo, E., Farley, C., Sharma, S.~K., Horton, K.,
  \& Granahan, J.~C. 2001, Journal of Geophysical Research-Planets, 106, 14595

\bibitem[{{Feaga} {et~al.}(2004){Feaga}, {McGrath}, {Feldman}, \&
  {Strobel}}]{Feaga2004}
{Feaga}, L.~M., {McGrath}, M.~A., {Feldman}, P.~D., \& {Strobel}, D.~F. 2004,
  \apj, 610, 1191

\bibitem[{{Feldman} {et~al.}(2001){Feldman}, {Ake}, {Berman}, {Moos}, {Sahnow},
  {Strobel}, {Weaver}, \& {Young}}]{Feldman2001}
{Feldman}, P.~D., {Ake}, T.~B., {Berman}, A.~F., {Moos}, H.~W., {Sahnow},
  D.~J., {Strobel}, D.~F., {Weaver}, H.~A., \& {Young}, P.~R. 2001, \apjl, 554,
  L123

\bibitem[{{Fink} \& {Sill}(1982)}]{fink1982}
{Fink}, U. \& {Sill}, G.~T. 1982, in IAU Colloq. 61: Comet Discoveries,
  Statistics, and Observational Selection, ed. L.~L. {Wilkening}, 164--202

\bibitem[{Grundy {et~al.}(2007)Grundy, Buratti, Cheng, Emery, Lunsford,
  McKinnon, Moore, Newman, Olkin, Reuter, Schenk, Spencer, Stern, Throop, \&
  Weaver}]{Grundy2007}
Grundy, W.~M., Buratti, B.~J., Cheng, A.~F., Emery, J.~P., Lunsford, A.,
  McKinnon, W.~B., Moore, J.~M., Newman, S.~F., Olkin, C.~B., Reuter, D.~C.,
  Schenk, P.~M., Spencer, J.~R., Stern, S.~A., Throop, H.~B., \& Weaver, H.~A.
  2007, Science, 318, 234

\bibitem[{Hand {et~al.}(2009)Hand, Chyba, Priscu, Carlson, Nealson, McKinnon,
  \& Khurana}]{Hand2009a}
Hand, K.~P., Chyba, C.~F., Priscu, J.~C., Carlson, R.~W., Nealson, K.~H.,
  McKinnon, W.~B., \& Khurana, K.~K. in 
  Europa, ed. R.~T. Pappalardo (Tucson, AZ: Univ. Arizona Press), 589

\bibitem[{Hudson \& Moore(2001)}]{Hudson2001}
Hudson, R.~L. \& Moore, M.~H. 2001, Journal of Geophysical Research-Planets,
  106, 33275

\bibitem[{{Kargel}(1991)}]{Kargel1991}
{Kargel}, J.~S. 1991, Icarus, 94, 368

\bibitem[{Kargel {et~al.}(2000)Kargel, Kaye, Head, Marion, Sassen, Crowley,
  Ballesteros, Grant, \& Hogenboom}]{Kargel2000}
Kargel, J.~S., Kaye, J.~Z., Head, J.~W., Marion, G.~M., Sassen, R., Crowley,
  J.~K., Ballesteros, O.~P., Grant, S.~A., \& Hogenboom, D.~L. 2000, Icarus,
  148, 226

\bibitem[{Kivelson {et~al.}(2000)Kivelson, Khurana, Russell, Volwerk, Walker,
  \& Zimmer}]{Kivelson2000}
Kivelson, M.~G., Khurana, K.~K., Russell, C.~T., Volwerk, M., Walker, R.~J., \&
  Zimmer, C. 2000, Science, 289, 1340

\bibitem[{{K{\"u}ppers} \& {Schneider}(2000)}]{Kuppers2000}
{K{\"u}ppers}, M. \& {Schneider}, N.~M. 2000, \grl, 27, 513

\bibitem[{Lane {et~al.}(1981)Lane, Nelson, \& Matson}]{Lane1981}
Lane, A.~L., Nelson, R.~M., \& Matson, D.~L. 1981, Nature, 292, 38

\bibitem[{Larkin {et~al.}(2003)Larkin, Quirrenbach, Krabbe, Aliado, Barczys,
  Brims, Canfield, Gasaway, LaFreniere, Magone, Skulason, Spencer, Sprayberry,
  \& Weiss}]{Larkin2003}
Larkin, J.~E., Quirrenbach, A., Krabbe, A., Aliado, T., Barczys, M., Brims, G.,
  Canfield, J., Gasaway, T., LaFreniere, D., Magone, N., Skulason, G., Spencer,
  M., Sprayberry, D., \& Weiss, J. 2003, Instrument Design and Performance for
  Optical/Infrared Ground-Based Telescopes, Pts 1-3, 4841, 1600

\bibitem[{McCord {et~al.}(2010)McCord, Hansen, Combe, \& Hayne}]{McCord2010}
McCord, T.~B., Hansen, G.~B., Combe, J.~P., \& Hayne, P. 2010, Icarus, 209, 639

\bibitem[{McCord {et~al.}(1998)McCord, Hansen, Fanale, Carlson, Matson,
  Johnson, Smythe, Crowley, Martin, Ocampo, Hibbitts, Granahan, \&
  Team}]{McCord1998a}
McCord, T.~B., Hansen, G.~B., Fanale, F.~P., Carlson, R.~W., Matson, D.~L.,
  Johnson, T.~V., Smythe, W.~D., Crowley, J.~K., Martin, P.~D., Ocampo, A.,
  Hibbitts, C.~A., Granahan, J.~C., \& Team, N. 1998, Science, 280, 1242

\bibitem[{McCord {et~al.}(2001)McCord, Hansen, \& Hibbitts}]{McCord2001}
McCord, T.~B., Hansen, G.~B., \& Hibbitts, C.~A. 2001, Science, 292, 1523

\bibitem[{McCord {et~al.}(1999)McCord, Hansen, Matson, Johnson, Crowley,
  Fanale, Carlson, Smythe, Martin, Hibbitts, Granahan, \& Ocampo}]{McCord1999}
McCord, T.~B., Hansen, G.~B., Matson, D.~L., Johnson, T.~V., Crowley, J.~K.,
  Fanale, F.~P., Carlson, R.~W., Smythe, W.~D., Martin, P.~D., Hibbitts, C.~A.,
  Granahan, J.~C., \& Ocampo, A. 1999, Journal of Geophysical Research-Planets,
  104, 11827

\bibitem[{McCord {et~al.}(2002)McCord, Teeter, Hansen, Sieger, \&
  Orlando}]{Mccord2002}
McCord, T.~B., Teeter, G., Hansen, G.~B., Sieger, M.~T., \& Orlando, T.~M.
  2002, Journal of Geophysical Research-Planets, 5004, 10.1029/2000JE001453, 107

\bibitem[{{McEwen}(1986)}]{mcewen1986}
{McEwen}, A.~S. 1986, \jgr, 91, 8077

\bibitem[{McKinnon \& Zolensky(2003)}]{McKinnon2003}
McKinnon, W.~B. \& Zolensky, M.~E. 2003, Astrobiology, 3, 879

\bibitem[{Moore {et~al.}(2007)Moore, Hudson, \& Carlson}]{Moore2007}
Moore, M.~H., Hudson, R.~L., \& Carlson, R.~W. 2007, Icarus, 189, 409

\bibitem[{Orlando {et~al.}(2005)Orlando, McCord, \& Grieves}]{Orlando2005}
Orlando, T.~M., McCord, T.~B., \& Grieves, G.~A. 2005, Icarus, 177, 528

\bibitem[{Paranicas {et~al.}(2001)Paranicas, Carlson, \&
  Johnson}]{Paranicas2001}
Paranicas, C., Carlson, R.~W., \& Johnson, R.~E. 2001, Geophysical Research
  Letters, 28, 673

\bibitem[{Paranicas {et~al.}(2002)Paranicas, Mauk, Ratliff, Cohen, \&
  Johnson}]{Paranicas2002}
Paranicas, C., Mauk, B.~H., Ratliff, J.~M., Cohen, C., \& Johnson, R.~E. 2002,
  Geophysical Research Letters, 29, 18

\bibitem[{Shirley {et~al.}(2010)Shirley, Dalton, Prockter, \&
  Kamp}]{Shirley2010}
Shirley, J.~H., Dalton, J.~B., Prockter, L.~M., \& Kamp, L.~W. 2010, Icarus,
  210, 358

\bibitem[{{Spencer} {et~al.}(2006){Spencer}, {Grundy}, {Dumas}, {Carlson},
  {McCord}, {Hansen}, \& {Terrile}}]{Spencer2006}
{Spencer}, J.~R., {Grundy}, W.~M., {Dumas}, C., {Carlson}, R.~W., {McCord},
  T.~B., {Hansen}, G.~B., \& {Terrile}, R.~J. 2006, Icarus, 182, 202

\bibitem[{{Strazzulla} {et~al.}(2007){Strazzulla}, {Baratta}, {Leto}, \&
  {Gomis}}]{Strazzulla2007}
{Strazzulla}, G., {Baratta}, G.~A., {Leto}, G., \& {Gomis}, O. 2007, Icarus,
  192, 623

\bibitem[{Zimmer {et~al.}(2000)Zimmer, Khurana, \& Kivelson}]{zimmer2000}
Zimmer, C., Khurana, K.~K., \& Kivelson, M.~G. 2000, Icarus, 147, 329

\bibitem[{{Zolotov}(2008)}]{Zolotov2008}
{Zolotov}, M.~Y. 2008, in Lunar and Planetary Inst. Technical Report, Vol.~39,
  Lunar and Planetary Institute Science Conference Abstracts, 2349

\end{thebibliography}
\end{document}